\documentclass[aps,prl,twocolumn,superscriptaddress,showpacs,amsmath,amssymb]{revtex4-1}
\usepackage{amsmath}
\usepackage{amsfonts}
\usepackage{amssymb}
\usepackage{graphicx}
\usepackage{color}

\usepackage{bm}
\usepackage{braket}
\usepackage{hyperref}

\begin{document}

\title{Anderson-Kitaev spin liquid}

\author{Masahiko G. Yamada}
\email[]{myamada@mp.es.osaka-u.ac.jp}
\affiliation{Department of Materials Engineering Science, Osaka University, Toyonaka 560-8531, Japan}
\affiliation{Institute for Solid State Physics, University of Tokyo, Kashiwa 277-8581, Japan}

\date{\today}

\begin{abstract}
The bond-disordered Kitaev model attracts much attention due to the
experimental relevance in $\alpha$-RuCl$_3$ and $A_3$LiIr$_2$O$_6$ ($A=$ H, D, Ag, \textit{etc.}).
Applying a magnetic field to break the time-reversal symmetry leads to
a strong modulation in mass terms for Dirac cones.  Because of the smallness
of the flux gap of the Kitaev model, a small bond disorder can have large
influence on itinerant Majorana fermions, and Majorana fermions will
be in the Anderson localization state immediately.  We call this immobile liquid state
Anderson-Kitaev liquid state with two localized Majorana fermions, one frozen
by gauge fluctuations and the other localized by disordered mass terms.
The quantization of the thermal Hall conductivity $\kappa/T$ disappears by a quantum Hall
transition induced by a small disorder, and $\kappa/T$ shows a rapid crossover
into the Anderson-Kitaev liquid with a negligible Hall current.  Especially,
the critical disorder strength $\delta J_{c1} \sim 0.05$ in the unit of the Kitaev interaction
would have many implications for the stability of Kitaev spin liquids.
\end{abstract}

\maketitle

\textit{Introduction}. ---
The Kitaev model~\cite{Kitaev2006} is one of the greatest examples of two-dimensional
(2D) solvable models of quantum spin liquids (QSLs)~\cite{Balents2010,Savary2017,Takagi2019},
especially in the perspective of spin-orbital-entangled
physics~\cite{Kitagawa2018,Yamada2018}.  This model has a bond-dependent anisotropic
interaction, which brings about exchange frustration and realizes gapped and gapless spin
liquid states depending on its parameters.  Amazingly, this interaction can be furnished
in materials with a strong spin-orbit coupling~\cite{Jackeli2009}.
Iridates and $\alpha$-RuCl$_3$ are prominent examples of candidate
materials for the Kitaev model~\cite{Singh2012,Plumb2014,Yamada2017},
but it is also known that these honeycomb materials cannot
fully be understood by the original (pure) Kitaev model~\cite{Chaloupka2010,Chaloupka2013}.
While other diagonal or offdiagonal interactions might be important in real materials~\cite{Song2016},
the importance of disorder has been ignored in these materials until
recently~\cite{Zschocke2015,Li2018,Knolle2019}.  Indeed, experiments in $A_3$LiIr$_2$O$_6$
($A=$ H, Ag, \textit{etc.}) show a universal scaling in the field dependence of
the heat capacity~\cite{Kitagawa2018,Bahrami2019}, which strongly suggests the existence of
disorder~\cite{Kimchi2018,Kimchi2018prx}.  The candidate
ground state must be disordered QSLs, and the absence of long-range order
can be attributed to the critical role of disorder.

In fact, the role of disorder in QSLs itself is a long-standing problem because of
the absence of a solvable model, except for limited cases~\cite{Yamada2019}.
We propose a disordered Kitaev model as a ``numerically''
solvable model for the disordered QSL, where we can treat the magnetic field effect
within the perturbation theory.  Thus, this study is not only a model investigation
for the disordered Kitaev materials like $A_3$LiIr$_2$O$_6$
($A=$ H, D, Ag, \textit{etc.})~\cite{Kitagawa2018,Bahrami2019,Geirhos2020},
but also a systematic examination of a numerically solvable disordered QSL, which would
be an attempt towards the universal understanding of various disordered QSLs.
Especially, since most QSLs are unsolvable, an unbiased study of
disordered QSLs was impossible in the previous method.
We invented a powerful numerical method based on kernel polynomial method (KPM)~\cite{Weisse2006}
to do a large-scale investigation ($O(10000)$ sites) for QSL.

Specifically, a Kitaev spin liquid (KSL)~\cite{Kitaev2006} is characterized by the
fractionalization of the spin into two types of Majorana fermions.  As such,
there is a possibility that an itinerant part of Majorana fermions will be
localized by the Anderson transition after introducing a quenched disorder.
This effect is strongest in 2D, but may be observable even in three-dimensional (3D)
generalizations~\cite{Obrien2016,Yamada2017xsl} (mobility edge).
These states with Majorana fermions in
an Anderson (weak) localization is named Anderson-Kitaev (AK) spin liquid,
or AK liquid in short.  We try to investigate the crossover
between KSL and AK liquid by the bond-disordered Kitaev model.

The pure Kitaev model is described by the following Hamiltonian:
\begin{align}
H_0 &= -J\sum_{\langle jk \rangle \in \gamma} \sigma_j^\gamma \sigma_k^\gamma,
\end{align}
where $\langle jk \rangle$ means a nearest-neighbor (NN) bond, and $J>0.$
$\gamma=x,$ $y,$ or $z$ is determined by a bond label.  This model is known to be solvable by
representing $\sigma_j^\gamma$ by Majorana fermions $ib_j^\gamma c_j.$  This representation
still works even if we introduce bond disorder as follows.
\begin{align}
    H_\textrm{bond} &= -\sum_{\langle jk \rangle \in \gamma} J_{jk} \sigma_j^\gamma \sigma_k^\gamma, \label{Hbond}
\end{align}
where $J_{jk}=J \pm \delta J$ is a bond-dependent hopping, and
$\delta J>0$ is the strength of bond disorder.  This model is still numerically solvable
if we can assume that the ground state is 0-flux when $\delta J$ is in the
perturbative regime.  Under this assumption, all the states with a pair of $\pi$-flux
vortices (vison) is assumed to be the ``first'' excited states from the ground state
flux sector.  This is how the perturbation theory works for this Kitaev model.
We employ Kitaev's trick to solve these Hamiltonians with an applied magnetic field~\cite{Kitaev2006}.

In this Letter, we simulate the bond-disordered Kitaev model to see a crossover
between KSL and AK liquid, especially from the topological transition in the thermal
Hall effect~\cite{Nasu2014,Nasu2017,Kasahara2018}.
We discovered that quantized thermal Hall effect is not as stable as
expected, and Majorana fermions are very easily localized by disorder.
Utilizing an approximation trick introduced by Kitaev, a large-scale calculation
up to $O(10000)$ sites is possible.  Important information for the Anderson transition
like density of states (DOS) has been calculated.

\begin{figure}
\centering
\includegraphics[width=8cm]{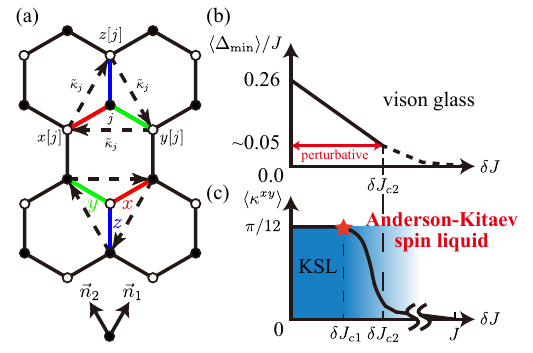}
\caption{Schematic phase diagram.  (a) Directional dependence of the bond interaction
and the NNN hopping arising for Majorana fermions under the magnetic field.
(b) $\Delta_\textrm{min} = \min (\Delta_\textrm{vison})$ versus disorder strength $\delta J.$
After the gap closing around $\delta J = \delta J_{c2},$ the flux sector becomes vison glass.
(c) $\kappa^{xy}$ versus disorder strength $\delta J.$  From $\delta J > \delta J_{c1},$
the crossover to the AK liquid is observed and $\kappa^{xy}$ finally reaches 0 around
$\delta J/J = 1.$}
\label{phase}
\end{figure}

\textit{Magnetic field effect}. ---
The Kitaev model on the honeycomb lattice can be defined from Fig.~\ref{phase}(a).
The bonds parallel to the red, green, and blue ones are $x$-, $y$-, and $z$-labeled bonds.
We first consider the pure Kitaev model with a magnetic field as follows.
\begin{align}
    H &= H_0 + V. \\
    V &= -\sum_j \left(h^x \sigma_j^x+h^y \sigma_j^y+h^z \sigma_j^z \right),
\end{align}
where $\vec{h} = (h^x,\,h^y,\,h^z)^t$ is an applied magnetic field.
We define a position operator $r_\alpha$ for the $\vec{n}_\alpha$-direction for $\alpha=1,\,2.$

It is well-known that $V_0$ can be treated by the third-order perturbation~\cite{Kitaev2006}.
The result after introducing itinerant Majorana fermions $c_j$ is
\begin{align}
    H_\textrm{eff} &= \frac{iJ}{2}\sum_{\langle jk \rangle} c_j c_k
    +\frac{i \tilde{\kappa}}{2}\sum_{\langle\!\langle jk \rangle\!\rangle} c_j c_k
    +(\textrm{four-fermion terms}). \\
    \tilde{\kappa} &= \frac{3h^x h^y h^z}{48 \alpha_0^2 J^2},
\end{align}
where $\alpha_0=0.262433$ in the thermodynamic limit for the 0-flux state, and
$\alpha_0 J$ is a vison gap in the uniform case.
The determination of the prefactor follows a mean-field solution~\cite{You2012}.
The direction of the bond $\langle\!\langle kl \rangle\!\rangle$ is defined clockwise
as shown in Fig.~\ref{phase}(a) around the site $j.$  A site connected by the $\gamma$-bond
from $j$ is called $\gamma[j]$ for $\gamma = x,$ $y,$ and $z,$ as shown in Fig.~\ref{phase}(a).
We define $\tilde{h} = h^x h^y h^z/48.$

\textit{Kitaev's trick}. ---
Next, let's include binary disorder as $H = H_\textrm{bond} + V.$
Following Kitaev~\cite{Kitaev2006}, we can always do perturbation from
any random $H_\textrm{bond}$ by a formula:
\begin{align}
    H_\textrm{eff}^{(3)} &= \Pi_0 V G_0^\prime(E_0) V G_0^\prime(E_0) V\Pi_0,
\end{align}
where $\Pi_0$ is a projection onto the ground state flux sector,
$G_0^\prime(E)$ is an unperturbed Green function constructed from $H_\textrm{bond}$
with the ground state flux sector excluded from the Hilbert space,
and $E_0$ is an initial energy.
Since $H_\textrm{bond}$ is solvable by Majorana fermions, it is in principle
possible to calculate $G_0^\prime(E)$ numerically to exhaust every term appearing
in the third order.
For example, a Green function for excited states is efficiently
obtained by the KPM~\cite{Weisse2006} numerically.
However, this strategy is surely overkill for our problem.

A much simpler solution is to use a trick introduced by Kitaev.  Though we still
need an $O(N^4)$ calculation cost to decide all terms by usual matrix diagonalization,
where $N$ is the number of sites,
there is no need for matrix exponentiation or integration.  Kitaev's trick is done
by replacing $G_0^\prime(E_0)$ by $-(1-\Pi_0)/\Delta_\textrm{vison},$ assuming that the virtual state
energy is constant determined just by a vison gap $\Delta_\textrm{vison}.$  This is a bold approximation
to simplify the problem drastically, but as we will see essential features, such as
the modulation of the mass term, can be captured even within Kitaev's approximation.

In this way, a typical third-order term is like the following:
\begin{align}
    H_\textrm{eff} &= \frac{i}{2}\sum_{\langle jk \rangle} J_{jk} c_j c_k
    +\frac{i}{2}\sum_{\langle\!\langle kl \rangle\!\rangle} \tilde{\kappa}_{kl} c_k c_l
    +\dots
\end{align}
where $\tilde{\kappa}_{kl}$ depends on the intermediate site $j$ in the third-order
perturbation process.  From $j,$ $\tilde{\kappa}$ can be calculated by replacing
$3/(\alpha_0 J)^2$ by $1/(\Delta_x \Delta_y) + 1/(\Delta_y \Delta_z) + 1/(\Delta_z \Delta_x),$
where $\Delta_\gamma$ is a vison gap for the bond between $j$ and $\gamma[j].$
\begin{equation}
    \tilde{\kappa}_{kl} = \tilde{\kappa}_j = \frac{\tilde{h}}{\Delta_x \Delta_y} + \frac{\tilde{h}}{\Delta_y \Delta_z} + \frac{\tilde{h}}{\Delta_z \Delta_x}.
\end{equation}
We note that three bonds have the same value of $\tilde{\kappa}_{kl}$ around $j.$
Thus, the disorder simply modulate the mass term of Dirac cones via random
NNN hoppings, and the problem is still solvable numerically.

In this case, four-fermion terms are short-ranged and irrelevant, so we have just ignored
them as we are only interested in the Hall conductivity in the $\vec{h} \to 0$ limit.
Though we will assume the ground state of $H_\textrm{bond}$ to be 0-flux in the following
discussions, the perturbation can be done from any flux configuration.
We note that a second-order perturbation in
$\vec{h}$ is ignored because it just renormalizes bond-dependent hoppings $J_{jk}$ and
does not break the time-reversal symmetry~\footnote{There is \textit{a priori} no way
to determine the ratio of the coefficients of the second- and third-order perturbations,
although we can always use a mean-field solution of the pure Kitaev model to estimate it~\cite{You2012}.}.

\begin{figure*}
\centering
\includegraphics[width=17.8cm]{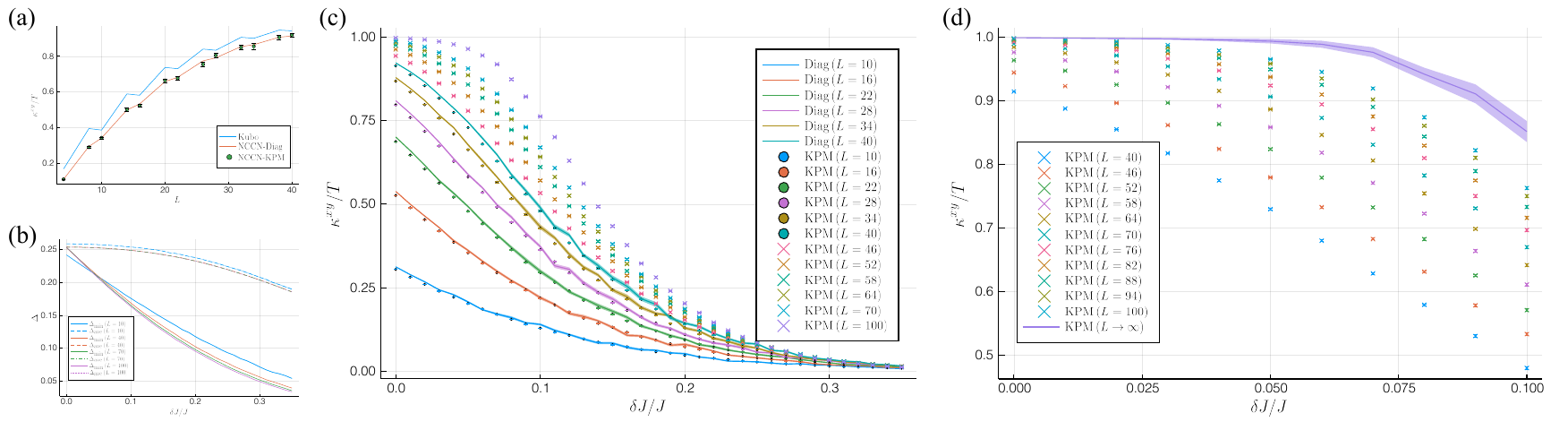}
\caption{(a) Kubo formula vs. NCCN.  NCCN-Diag means that the Chern number
is calculated by diagonalization, while NCCN-KPM means that the Chern number is
calculated by KPM with $M^\prime=512$ and $R=100.$  In order to put errorbars,
random vectors are chosen to be Haar-random.  Only $L \mod 6 = 2,\,4$ is
plotted.  The ordinary periodic boundary condition is used.
(b) Mean and minimum values of flux gaps calculated by KPM. The errorbar is smaller than
the line width and only plotted for $L=10.$  KPBC is used
(c) NCCN calculated by diagonalization (solid lines) and KPM (scatter plots).
$N_\textrm{sample} =24$ is used for the diagonalization. For KPM we used $R=24,$
and $N_\textrm{sample} = 360.$  KPBC is used.
(d) NCCN calculated by KPM and the value extracted for $L \to \infty.$
The margin of error at 5\% significance level is used for the ribbon for the extrapolation.}
\label{ed}
\end{figure*}

\textit{Thermal conductivity}. ---
We only consider zero temperature and ignore thermal flux fluctuations above the 0-flux sector.
Lieb's theorem~\cite{Lieb1994} no longer applies, but we can expect it to
be applicable on average.  Anyway, the calculation is relevant only in the regime
where the flux gap is not closed by thermal fluctuation or bond disorder
($\delta J < \delta J_{c2}$ in Fig.~\ref{phase}(b)-(c)).

We employed Kitaev's trick to calculate a Majorana spectrum with an external
magnetic field for each quenched bond disorder.  From this, we can compute
an in-plain thermal Hall conductivity $\kappa^{xy}(T),$ especially a behavior
of $\kappa^{xy} / T$ at $T \to 0.$  Here $xy$ does not coincide with the Cartesian
axis but means a transverse component of the thermal conductivity.
A Kubo formula for $\kappa^{xy}$ at zero temperature is reduced to the
generalized Thouless-Kohmoto-Nightingale-den Nijs (TKNN) formula~\cite{Thouless1982} for noninteracting
Majorana Hamiltonians~\footnote{The generalized formula without a translation symmetry
is originally discussed by Kitaev~\cite{Kitaev2006} using a flow of unitary matrices
and extended by Kapustin and Spodyneiko~\cite{Kapustin2020}.}:
\begin{align}
    \kappa^{xy} &= \frac{\hbar \pi^2 k_B^2 T}{6L^2}
    \sum_{m,n} \vartheta(-\varepsilon_m)\frac{2 \textrm{Im}[\braket{m|v_1|n}\braket{n|v_2|m}]}{(\varepsilon_m - \varepsilon_n)^2} ,
\end{align}
where $m$ and $n$ label eigenvalues of $\mathcal{H},$ $\varepsilon_m$ and $\varepsilon_n,$
corresponding to eigenstates $\ket{m}$ and $\ket{n},$ respectively~\cite{Nomura2012,Sumiyoshi2013}.
$\vartheta(x)$ is a Heaviside theta and $v_\alpha = i[\mathcal{H}, r_\alpha]/\hbar$ is a
velocity operator along the $\alpha$-direction.
This Kubo-TKNN formula~\footnote{A gravitomagnetic term should be added to derive
this formula.} is nothing but a real-space formulation of the Chern number
calculation.

We can alternatively use the so-called noncommutative
Chern number (NCCN)~\cite{Prodan2010}, which is defined by a spectral projector for occupied free fermions.
This formula is advantageous because it is proven to become integer after disorder
average with some conditions, whereas it only makes sense in the thermodynamic limit.
\begin{equation}
    \textrm{Ch} = -\frac{2\pi i}{L^2}\,\textrm{tr}\left\{P_F \left[[r_1, P_F], [r_2, P_F]\right]\right\},
\end{equation}
where $P_F= \sum_n \vartheta(-\varepsilon_n) \ket{n}\bra{n}$ is a spectral projector.
These two formulae must agree in the thermodynamic limit by a well-known relation
$\kappa^{xy}/T = \pi k_B^2\mathrm{Ch}/(12 \hbar)$ for Majoranas.
The finite-size effect is suppressed exponentially by an artificial $k$-space quantization
of a size $L \times L$ and by replacing the commutator~\cite{Prodan2010} as follows:
\begin{equation}
    -i[r_\alpha, P_F] \mapsto \sum_{q = -Q}^{Q} c_q e^{-iq\Delta r_\alpha} P_F e^{iq\Delta r_\alpha},
\end{equation}
where $\Delta = 2\pi/L,$ $c_0=0$ and $c_q=-c_{-q}$ are determined to hold
$x-\sum_{q = -L/2}^{L/2}c_q e^{iq\Delta x}=O(\Delta^L),$ and $Q \leq L/2.$
When $Q = L/2,$ this formula exponentially converges to the thermodynamic limit
with a self-converging property.
Thus, we can expect that these two methods may agree with a large $L,$
while the Hall conductivity and the Chern number are \textit{a priori}
different quantities.  We note that there are other ways to detect the topological
nontriviality~\cite{DeNittis2016,Akagi2017,Katsura2018}.

After taking an average of $\kappa^{xy} / T$
over a number of disorder configurations, we plot a physical thermal Hall conductivity
as a function of $\delta J.$  The error bar is estimated from a statistical
deviation.  From now on we set $\hbar = k_B = J = 1.$

\textit{Numerical results}. ---
We first note that, since we only include the third-order perturbation, the results
here are not simply comparable with experiments.  However, it was proposed that
the contribution from $h^x h^y h^z$ can be picked up by applying an inplane magnetic
field~\cite{Yokoi2020}, so we only take an odd component under every sign change
($h^x \mapsto -h^x,$ $h^y \mapsto -h^y,$ and $h^z \mapsto -h^z$)
of the three components of $\vec{h}$ from total $\kappa^{xy}.$
From now on we denote $\kappa^{xy}$ as an odd component under every sign change
and ignore other components.

The approximate correspondence between the Kubo formula and NCCN is confirmed
for the pure Kitaev model [see Fig.~\ref{ed}(a)].  We note that Haar-random
vectors used in this calculation show large errorbars and are not used in the
following as described in Supplemental Material (SM)~\cite{SM}.
From here we will prefer the NCCN because we can use
the KPM to approximate the spectral projector $P_F$ to avoid
the diagonalization~\footnote{Application of KPM to the Kubo formula requires efforts~\cite{Garcia2015}.}.
We fixed $Q = 15$ for $L > 30$ because otherwise the calculation cost becomes $O(N^3).$
KPM can reproduce the vison gap approximately and at most reduce the computational
cost to $O(N)$ with a truncation~\cite{Furukawa2004,Ishizuka2012,Ishizuka2013}.
However, later we found that the truncation cause
a problem in our simulation, and thus we used the $O(N^2)$ algorithm~\cite{Weisse2006,Weisse2009,Mishchenko2017}.

Next, we would move on to a large-scale calculation by Kitaev's trick.
From now on, $\kappa^{xy}$ is always calculated through NCCN.
We only take (Kitaev's) $L \times L$ periodic boundary condition (for spins) from $L=10,$ where
the vison gap gets close to the thermodynamic limit.  As long as we are interested in
the topological property the $\vec{h} \to 0$ limit does not have to be taken.
We set $h^x = h^y = h^z \equiv h = \Delta_\textrm{min},$ where $\Delta_\textrm{min}$
is the minimum vison gap as a vison gap has spatial dependence on each bond,
for simplicity~\footnote{In reality, $h/\Delta_\textrm{min}$ must be smaller than unity,
but this suffers from the finite-size effect.}.  In order to reduce the finite-size effect,
we adopt Kitaev's torus basis where the finite-size effects cancel out,
which is defined from a torus basis $(L \vec{n}_1, L \vec{n}_2 + \vec{n}_1)$~\cite{Kitaev2006}.
We call it Kitaev's periodic boundary condition (KPBC) for simplicity.
The NCCN formula for KPBC has to be modified as described in SM~\cite{SM}.
This arbitrary choice of boundary conditions does not matter in the thermodynamic limit.
The averaged $\langle \kappa^{xy} \rangle / T$ for $T \to 0$ is shown as
a function of $\delta J,$ and drops rapidly to 0 from the quantized value
as the disorder strength $\delta J$ grows.  From here $\langle \kappa^{xy} \rangle / T$
is plotted in the unit of a quantum $\pi/12.$
We used $R=24$ vectors to approximate the trace~\cite{Varjas2020}.

The mean and minimum value of vison gaps are plotted for each $\delta J$ in Fig.~\ref{ed}(b).
When $\delta J > \delta J_{c2} \sim 0.3,$ the vison gap approaches 0 for some plaquette, and
the 0-flux ground state is destabilized.  From here, the perturbation from the
0-flux sector cannot be justified.  Moreover, after the gap closing, some flux sectors
get almost degenerate and the first-order perturbation in $\vec{h}$ now becomes relevant.
Beyond this point, a quantized thermal Hall current is no longer a well-defined notion.
Flux excitations and (itinerant) Majorana fermions are not separable, and the discussion
based only on free Majorana fermions breaks down.

When $\delta J \ll \delta J_{c2},$ the calculation by Kitaev's trick can be justified.
Fig.~\ref{ed}(c) shows NCCN calculated by diagonalization (line plot) and KPM (scatter plot).
These two methods agree well.  From the data of KPM we extrapolated the thermodynamic limit.
The finite-size data are fit by exponential functions, and extracted the converged value
for $L \to \infty.$  The extrapolation is plotted in Fig.~\ref{ed}(d) and the thermodynamic
limit is shown in a line plot with a ribbon.  Around $\delta J = \delta J_{c1} = 0.05,$
NCCN deviates from unity, which suggests the existence of the topological transition
into the gapless phase.
For the calculations we took $N_\textrm{sample} = 360$ quenched disorder samples
and used $M=1024,$ and $R=24,$ where $M$ is the expansion order of KPM.

\begin{figure}
\centering
\includegraphics[width=8.6cm]{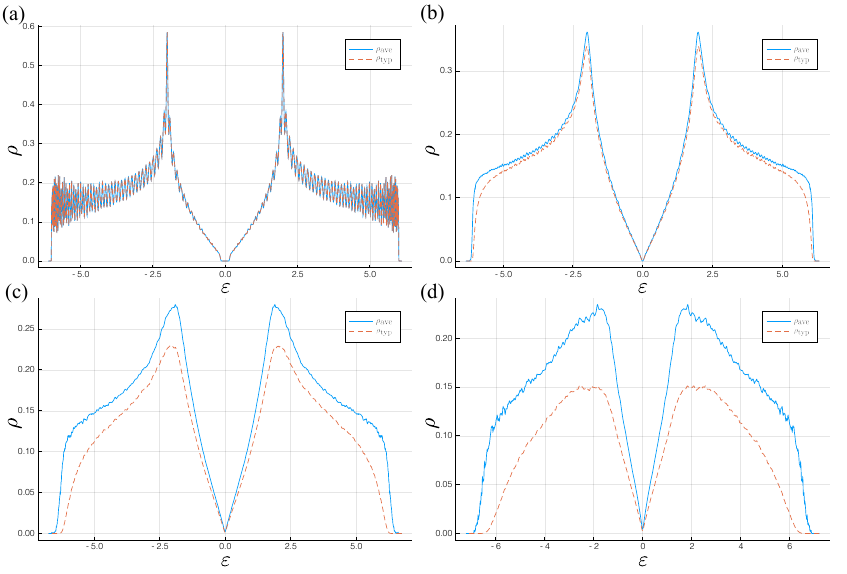}
\caption{Arithmetic and geometric means of LDOS.  (a) $\delta J = 0.0.$
(b) $\delta J = 0.1.$  (c) $\delta J = 0.2.$  (d) $\delta J = 0.3.$
For every figure $L=100,$ $R=24,$ and $N_\textrm{sample} = 360.$}
\label{ldos}
\end{figure}

\textit{Localization of Majorana fermions}. ---
When $\delta J \ll \delta J_{c2},$ free Majorana fermions are only
relevant low-energy excitations, and we can use many tools of free fermions
to discuss properties of the transition, such as DOS and a localization length.
DOS around the ground state can be measured from the information of the 0-flux sector.
As is often the case, we only calculated local density of states (LDOS), instead.
The nonlocality of Majorana fermions does not matter as averaged LDOS approximates DOS well
enough. Both of the quantities are easily computed using KPM, and LDOS is enough for our
purpose.  The ratio of the arithmetic and geometric means of LDOS also works as
the order parameter of an Anderson transition instead of the localization length.
From the gapped Dirac spectrum [see Fig.~\ref{ldos}(a)] the LDOS becomes gapless
as the disorder strength increases.  In the gapless region, DOS behaves linearly
around $\varepsilon = 0$ [see Fig.~\ref{ldos}(b)-(d)].
The localization in Fig.~\ref{ldos}(b)-(d) is clear from the discrepancy between
the arithmetic and geometric averages of LDOS.  Details are included in SM~\cite{SM}.

\textit{Discussions}. ---
Though we only did a finite-size calculation, the transition between KSL and
AK liquid was well-observed and the schematic phase diagram in Fig.~\ref{phase}(c)
was confirmed.  From the extrapolation,
$\delta J_{c1}$ is very small and $\delta J_{c1}/J \sim 0.05.$
This fragility may be related to the long-range correlation
in the mass term disorder~\cite{Fedorenko2012},
and reflects the nonlocality of the definition of Majorana fermions.
We note that the vortex disorder is known to be relevant, so the introduction of
random vortices may change the universality~\cite{Bocquet2000}.
After the transition the V-shaped behavior of DOS completely agrees with an observed
linear low-energy DOS for H$_3$LiIr$_2$O$_6$ with an applied magnetic field~\cite{Kitagawa2018}.

The fragility of the quantization has many implications to experiments.  Disorder always exists
in real materials, especially in any 2D layered system, and even in clean samples of
$\alpha$-RuCl$_3$ stacking faults must exist~\cite{Yamauchi2018}.  Thus, the situation is quite
similar to that of the fractional quantum Hall effect (FQHE).  The observation
of FQHE requires a really clean sample, and the recently observed quantized thermal
Hall current of FQHE is more sensitive to disorder~\cite{Banerjee2018}.
The sensitivity also resembles unconventional superconductors~\cite{Ngampruetikorn2020}.
It might be universal in strongly correlated systems.
Thus, we need to reconsider the importance of cleanness for the topological order in general.
Last but not least, we fixed $h/\Delta_\textrm{min} = 1.0$ for simplicity,
so it is necessary to check another parameter region for comparison.

\begin{acknowledgments}
We thank Y.~Akagi, S.~Fujimoto, H.~Ishizuka, G.~Jackeli, H.~Katsura, I.~Kimchi, Y.~Matsumoto, T.~Matsushita,
T.~Morimoto, N.~B.~Perkins, Y.~Tada, D.~Takikawa, K.~Totsuka, and S.~M.~Winter.
M.G.Y. thanks G.~Chen for suggesting a new title.
M.G.Y. is supported by the Materials Education program for the future leaders in Research, Industry, and Technology (MERIT), and by JSPS.
This work was supported by JST CREST Grant Number JPMJCR19T5, Japan,
and by JSPS KAKENHI Grant Numbers JP17J05736 and JP17K14333.
This research was supported in part by the National Science Foundation under Grant No. NSF PHY-1748958.
The computation in this work has been done using the facilities of the Supercomputer Center, the Institute for Solid State Physics, the University of Tokyo.
\end{acknowledgments}

\bibliography{paper}

\end{document}


\title{Supplemental Material for ``Anderson-Kitaev spin liquid''}

\author{Masahiko G. Yamada}
\email[]{myamada@mp.es.osaka-u.ac.jp}
\affiliation{Department of Materials Engineering Science, Osaka University, Toyonaka 560-8531, Japan}
\affiliation{Institute for Solid State Physics, University of Tokyo, Kashiwa 277-8581, Japan}

\maketitle

\section{Kernel polynomial method}

We would like to introduce the kernel polynomial method (KPM)~\cite{Weisse2006,Weisse2009,Mishchenko2017}.
For this approximation, We wrote a program in the Julia 1.3.1 language.

First, let's consider a Majorana Hamiltonian with the following form.
\begin{equation}
    H = \frac{1}{4}\sum_{j,k}\mathcal{H}_{jk} c_j c_k,
\end{equation}
where $\mathcal{H}$ is a Hermitian matrix.  For Majorana fermions $c_j,$ $\mathcal{H}$ has a form
$\mathcal{H} = iA,$ where $A$ is a real skew-symmetric matrix.  From now on, we assume $\mathcal{H}$
to be the ones considered in the main text, either with or without a magnetic field.
The eigenvalues of the $N \times N$ matrix $\mathcal{H}$ is denoted by $E_k$
with $k = 1,\dots,N.$

A Green function can be expanded by a Chebyshev polynomial $T_m(x)$ as follows.
\begin{align}
    G_{jj}(E+i\varepsilon) &= i\frac{\tilde{\mu}_0 + 2\sum_{m=1}^M \tilde{\mu}_m \exp[-im \arccos(E/s)]}{\sqrt{s^2 - E^2}}. \\
    \tilde{\mu}_m &= g_m \bra{j} T_m(\mathcal{H}/s) \ket{j}. \\
    g_m &= \frac{\sinh[\lambda (1-m/M)]}{\sinh \lambda},
\end{align}
where $\lambda=4.0$ was used in the Lorentz kernel $g_m.$  $M$ is the expansion order and
$m = 0,\dots,M-1.$  $\varepsilon$ is a small parameter which goes to 0 when $M \to \infty.$
The scaling factor $s$ is necessary so that the spectrum of the Hamiltonian $\mathcal{H}/s$
falls within the domain of the Chebyshev polynomials $[-1,1].$
We note that this expression is for diagonal components, but almost the same is true for
offdiagonal components.
From the connectivity of the Hamiltonian we can set $s = 6(J+\delta J)$ without a magnetic
field, but it is more convenient to use Arpack.jl or ArnoldiMethod.jl to compute the maximum
absolute value $E_\textrm{max}$ of eigenvalues, and set
$s = E_\textrm{max} + 0.1.$

Elements of Chebyshev moments $T_m(\mathcal{H}/s)$ can be computed recursively
by using $T_m(x) = 2xT_{m-1}(x) - T_{m-2}(x)$ and $T_{2m+i}(x) = 2T_m(x) T_{m+i}(x) - T_i(x)$
for $i=0,\,1.$  The total $O(N^2)$ cost is required to compute all the necessary elements.

From the expanded Green function, we can compute the energy change by the local modification
of the Hamiltonian $\mathcal{H} \to \mathcal{H} + \delta \mathcal{H}.$  We define
\begin{equation}
    \mathcal{D}(E) = \det[1+G(E)\delta \mathcal{H}].
\end{equation}
By extending this function to a complex number, the energy difference, \textit{i.e.}
vison gap $\Delta,$ can be computed as follows.
\begin{equation}
    \Delta = \frac{1}{\pi} \int_{-\infty}^\infty F(E)\lim_{\varepsilon \to 0}\textrm{Im}\,\log [\mathcal{D}(E+i\varepsilon)]dE,
\end{equation}
where $F(E)$ is a Fermi-Majorana function at zero temperature.
\begin{equation}
    F(E) = -\lim_{\beta \to \infty}\frac{1}{2}\tanh\frac{\beta E}{2} = \vartheta(-E) - \frac{1}{2}.
\end{equation}

The evaluation of the integral in the Green function requires fast
Fourier transformation (FFT) or discrete cosine transformation (DCT)~\cite{Weisse2006}.
Fortunately, Julia has a wrapper for
FFTW~\footnote{Frigo, M., and S. G. Johnson, 2005b, FFTW fast fourier transform library.}
called FFTW.jl.  Using this, the integral is reduced to a discrete weighted sum of the
Fermi(-Majorana) function evaluated at some specific points.  FFT (type-III DCT for
the real diagonal part) decreases the computational
cost drastically for evaluating the Chebyshev polynomials.  The discretization size
$\tilde{M}$ was set $\tilde{M} = 2M$ for simplicity.  We tried $M=64,\,128,\,256,\,512,\,1024,$
and $2048.$  We found that $M = 1024$ has the best performance for our purpose,
where the error is always about $0.01J.$

As for the estimation of a noncommutative Chern number (NCCN)~\cite{Prodan2010},
it is better to expand the Fermi function directly instead of an FFT scheme.
The spectral projector $P_F$ can be written as:
\begin{equation}
    P_F = -\frac{1}{2\pi i}\oint_{\mathcal{C}} \frac{1}{\mathcal{H} -E} dE,
\end{equation}
where $\mathcal{C}$ is a contour which encloses every negative eigenvalue of $\mathcal{H}.$
The integrant is nothing but a Green function, so this can be expanded by KPM.

It is better to use the following Fermi function instead of approximating a spectral
projector directly.
\begin{equation}
    P_F^\textrm{eff} = F(\mathcal{H}/s) = \vartheta(-\mathcal{H}/s) - \frac{1}{2}.
\end{equation}
Due to the particle-hole symmetry, this deformation also gives a correct Chern number.
This expression can again be expanded by Chebyshev polynomials.
Especially, the Fermi function has been expanded as
\begin{align} 
    f_m &= \int_{-1}^1 \frac{dx}{\pi \sqrt{1-x^2}} T_m(x) F(x). \\
    F(x) &= f_0 + 2\sum_{m=1}^{M^\prime - 1} f_m T_m(x),
\end{align}
where $M^\prime$ is a cutoff of the expansion for NCCN~\cite{Varjas2020}.
We used the Jackson kernel for the Chern number:
\begin{align} 
    g_m^\prime &= \frac{(M^\prime - m + 1) \cos\frac{\pi m}{M^\prime + 1} + \sin\frac{\pi m}{M^\prime + 1} \cot\frac{\pi}{M^\prime + 1}}{M^\prime + 1},
\end{align}
where $m = 0,\dots,M^\prime - 1.$  We here used $M^\prime = 512,$ instead.  Thus,
\begin{equation}
    P_F^\textrm{eff} \sim g_0^\prime f_0 T_0(\mathcal{H}) + 2\sum_{m=1}^{M^\prime - 1} g_m^\prime f_m T_m(\mathcal{H}).
\end{equation}
We will again use $M=1024$ for local density of states (LDOS) later.

\section{$O(N)$ approximation of the Chern number}

\begin{figure}
\centering
\includegraphics[width=8.6cm]{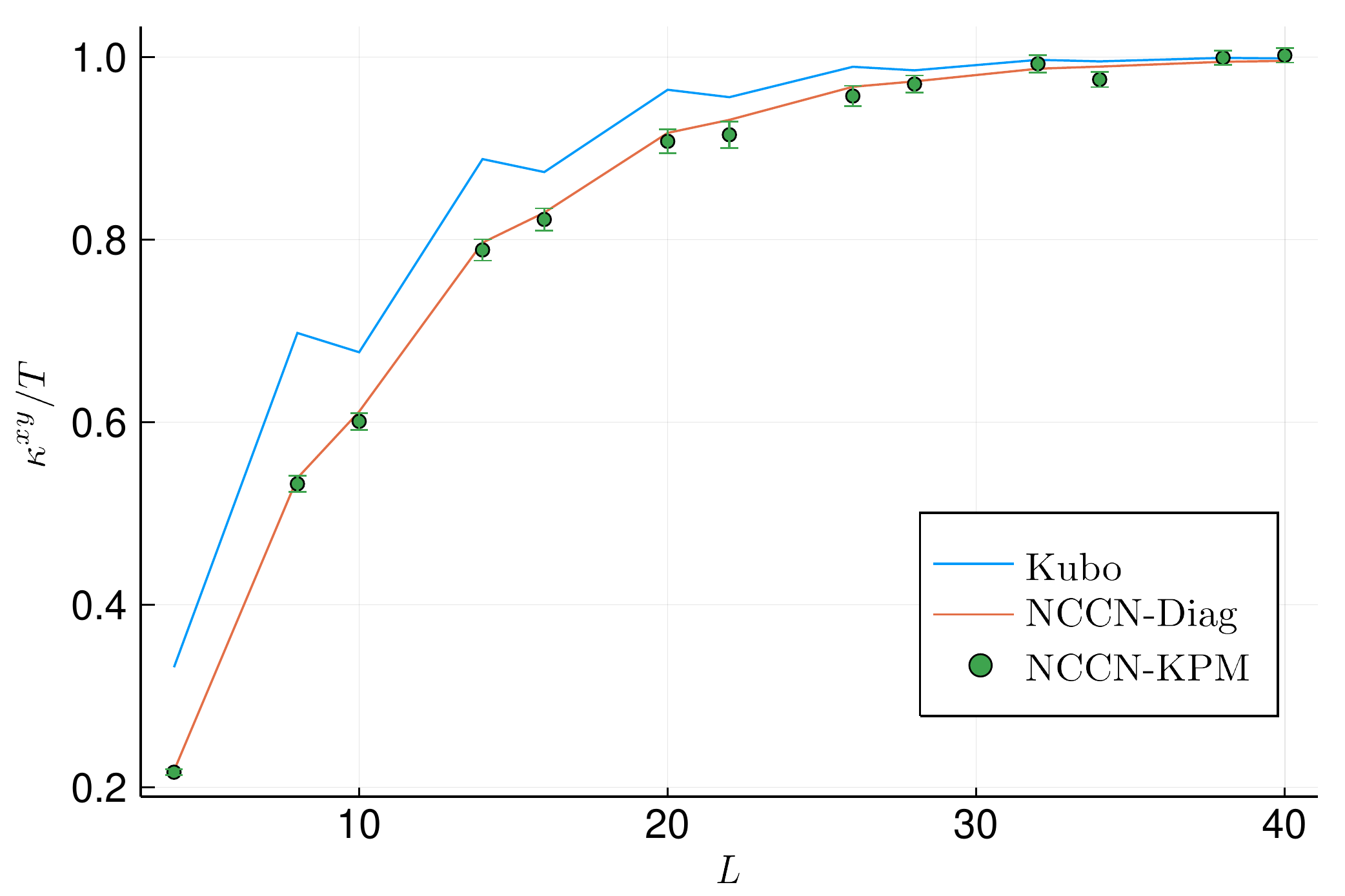}
\caption{Convergence of the Kubo formula and NCCN.  NCCN-Diag means that the Chern number
is calculated by diagonalization, while NCCN-KPM means that the Chern number is
calculated by KPM with $M^\prime=512$ and $R=100.$  In order to put errorbars,
random vectors are chosen to be Haar-random.  Differently from the main text,
$h/\Delta_\textrm{min}$ is set $\sqrt[3]{2}.$
The ordinary periodic boundary condition is used.}
\label{comp}
\end{figure}

If the method above is directly applied, the estimation of $P_F$ still requires $O(N^2).$
We propose a stochastic method
to evaluate the Chern number based on the randomized algorithm to estimate a trace~\cite{Weisse2006}.
By picking up $R$ normalized random vector $\ket{r},$ the trace is approximated as follows.
\begin{equation}
    \frac{1}{N}\textrm{Tr}\,B \approx \frac{1}{R}\braket{r|B|r},
\end{equation}
where $B$ is some $N \times N$ matrix.
We can use a Haar-random vector, a $Z_2$-random vector, \textit{etc}.
In Fig.~\ref{comp}, $R=100$ Haar-random vectors are used.
Fig.~\ref{comp} shows the convergence of the Kubo formula
and NCCN better than in the main text.  An unphysically large magnetic field
of $h/\Delta_\textrm{min} = \sqrt[3]{2}$ is applied.

However, it is actually better
to use a site basis $\ket{i}$ for a random site $i$ to take a trace.
Actually, in the thermodynamic limit NCCN is expected to be independent of $i$~\cite{Prodan2010}.
$R=24$ is enough for our purpose, and the result gets accurate as the system
size increases.

In this $O(N)$ method, $q = 1,\dots,Q$ terms are all expanded.
In the large scale we fixed $Q = 15,$ so we need to evaluate $Q^2 = 225$ terms at the same time.
We note that terms with a negative $q$ can easily be computed from the positive ones.

Instead of calculating every element of $P_F,$ we can now estimate
$P_F \ket{\alpha}$ for any (sparse or dense) vector $\ket{\alpha}$ with an $O(NM)$ cost.
Since $r_1$ and $r_2$ are both diagonal in the original basis, the trace can
be computed by sequential estimations of $P_F \ket{\alpha}$ after decomposing
the summation over $q.$  Eventually, the calculation costs becomes $O(NMQR + N Q^2 R),$
and it scales as $O(N)$ as long as $M,$ $Q,$ and $R$ are kept constant.
We note that the preparation of the look-up tables has been ignored.

\section{Random-walk $O(N)$ approximation for the Chebyshev series}

\begin{figure}
\centering
\includegraphics[width=8.6cm]{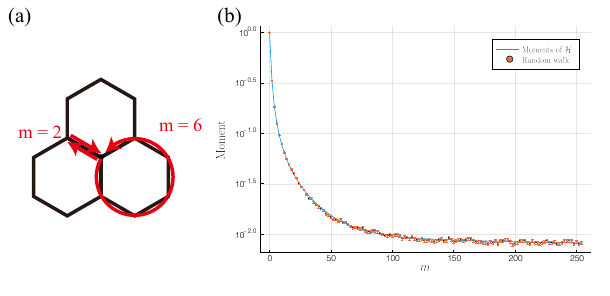}
\caption{(a) Examples of moment counting in the weighted random walk.
(b) Comparison between the exact moment of the Hamiltonian and the random-walk estimation.
$L = 10,$ $J = 1.0,$ $\delta J = 0.01.$  KPBC is used without a magnetic field.
For the random walk, $N_\textrm{walker} = 100000,$ $a = 6.0,$ and moments were
plotted until $M=256.$}
\label{rand}
\end{figure}

Though in the main text we only used a conventional $O(N^2)$ method for calculating
the Chebyshev series, there exists a stochastic algorithm which can approximate
the series with a computational cost of $O(N)$ \textit{without any truncation.}
This algorithm can work as a Markov
chain simulation of the matrix product process and the probability that random walkers
go back to the original site gives each (diagonal) element of the moment of the Hamiltonian.
Thus, we can use this weighted random walk to obtain every (diagonal or offdiagonal) element
of the moments, and we will quickly demonstrate the algorithm for diagonal cases,
and show that all the diagonal parts can be obtained in $O(N)$ time.
This is nothing but an importance sampling among every contour which goes from the
site $i$ to the same site $i,$ and the weight stays constant, \textit{i.e.}
the importance sampling goes well, when the disorder strength is small enough.
Examples are shown in Fig.~\ref{rand}(a).

The algorithm is very simple. The number of walkers $N_\textrm{walker}$ should be kept
constant.
\begin{algorithm}[H]
\caption{Weighted Random Walk}
\begin{algorithmic}[1]
    \State{set $a$ as an arbitrary scaling factor}
    \For{$m \in \{1,\dots,M-1\}$}
        \State{$h_m \gets 0$}
    \EndFor
    \For{$j \in \{1,\dots,N_\textrm{walker}\}$}
        \State{$p_j \gets i$}
        \State{$w_j \gets 1.0$}
    \EndFor
    \For{$m \in \{1,\dots,M-1\}$}
        \For{$j \in \{1,\dots,N_\textrm{walker}\}$}
            \State{$\nu(p_j) \gets$ nearest neighbor sites of $p_j$}
            \State{$\omega(p_j) \gets$ hopping amplitudes from $p_j$}
            \State{$w_j \gets w_j \times$(the sum of $\omega(p_j)$)/$a$}
            \State{sample $p_j$ from $\nu(p_j)$ with weights $\omega(p_j)$}
            \If{$x_j=i$}
                \State{$h_m \gets h_m + w_j$}
             \EndIf
        \EndFor
    \EndFor
    \State{\Return{$h_m$}}
\end{algorithmic}
\end{algorithm}
The returned histogram $h_m$ directly gives an estimation of the moment of
Hamiltonian $\braket{i|\mathcal{H}^m|i}$ with a scaling factor $a^m.$
Thus, if we need an $O(N)$ cost to exhaust every diagonal component.
We note the offdiagonal component can be computed by taking a histogram at site $j.$
In the weak-disorder regime, the growth of weight can be kept almost unity, so
the histogram itself can be approximated by a Poisson distribution.  Thus, we
put an errorbar for the number of events $\tilde{N}$ as $\sqrt{\tilde{N}}.$
The calculation is compared with the exact results in Fig.~\ref{rand}(b).

One of the advantages of this method is that by replacing $\mathcal{H}$ by
$\mathcal{H} + sI,$ where $I$ is an identity matrix, the transition matrix
can always be made positive-definite.  The convergence is guaranteed by the
Perron-Frobenius theorem, and the relaxation time is determined from the difference
between the first and second largest eigenvalues.  Thus, after the relaxation,
all the moments can be replaced by the equilibrium value and we can reconstruct
the Chebyshev series until $M \to \infty.$

However, the prefactor of the cost of this algorithm is very large, so we rather
prefer the $O(N^2)$ deterministic algorithm, instead, in this work.  This $O(N)$
algorithm is advantageous only in a very large scale $N \sim 10^6$ and we would
pursue its usefulness in the future.  Especially, in the strongly disordered case,
the weight $w_j$ should be balanced by a node-weighted algorithm.

In the flux sector except for the 0-flux one, this algorithm strongly suffers from
a ``sign problem''.  Contours enclosing a $\pi$ flux will be affected by this sign problem.
Dealing with a general flux based on this algorithm is also an important future problem.
We note that this method can be combined with
a truncation by excluding walkers which go too far from the origin.

\section{Kitaev's periodic boundary condition}

In our method, the Chern number calculation assumes infinite supercells of the periodic
boundary condition, and also the unitarity of the discrete Fourier transformation.
Thus, if we employ Kitaev's periodic boundary condition (KPBC)~\cite{Kitaev2006},
we must do a linear
transformation to retain the periodicity of plane waves.  The coordinate
transformation is necessary from the usual crystallographic coordinate $(r_1,\,r_2)$
to $(r_1 - r_2 / L,\,r_2).$  In this new basis, the computation of NCCN is possible.
Apparently, this basis change does not affect the thermodynamic limit.

\section{Evaluation of local density of states}

LDOS at site $i$ is defined as follows~\cite{Weisse2006}.
\begin{equation}
    \rho_i(E) = \sum_{k=1}^N |\braket{i|k}|^2 \delta(E - E_k),
\end{equation}
where $\ket{k}$ is an eigenstate of $\mathcal{H}$ corresponding to an eigenvalue $E_k.$
Apparently, the average over every $i$ would be the density of states (DOS).
As long as the system is uniform, LDOS coincides with DOS.

LDOS can be calculated through KPM.  The expansion is like
\begin{align}
    \mu_m &= \int_{-1}^1 \rho_i(E) T_m(E) dE = \sum_k \braket{i|T_m(\mathcal{H})|k}\braket{k|i} \nonumber \\
    &= \braket{i|T_m(\mathcal{H})|i}.
\end{align}
We note $M=1024$ is used for LDOS.
Usually, $i$ is randomly chosen in the same way as the Chern number calculation, and
the arithmetic mean of $\rho_i(E)$ is denoted by $\rho_\textrm{ave}.$
Its geometric mean $\rho_\textrm{typ}$ over a number of samples is also important.
\begin{equation}
    \rho_\textrm{typ} = \exp[\langle\!\langle \log(\rho_i(E)) \rangle\!\rangle],
\end{equation}
where the disorder average is taken in the expression.

The reason why the difference in the arithmetic and geometric averages of LDOS captures
the localization is as follows.  If the localization happens, the LDOS strongly depends
on the site, and LDOS at site $i$ deviates from the true DOS (averaged LDOS).
Thus, the typical LDOS ($\rho_\textrm{typ}$) deviates from the averaged LDOS
($\rho_\textrm{ave}$), and it signals the localization transition.
Though we have plotted LDOS for
one-body Majorana Hamiltonians, which is not necessarily the same as the many-body
(L)DOS, the results can always be transformed into the many-body language,
which will be done in the future work.

As for another signal, the localization length is also calculable using
the real-space Kubo-Greenwood formalism~\cite{Fan2014},
but we would leave it as future work because LDOS already worked as the proof of the
localization.

\section{Future direction}

In reality, a problem of solving the disordered Kitaev model is
many-body localization (MBL)~\cite{Theveniaut2019}
because there is a four-fermion term~\cite{Kitaev2006}.
The effect of a finite interaction should be
studied in the future to find out some MBL criterion.  The problem itself resembles
physics of the Sachdev-Ye-Kitaev model and its MBL~\cite{Maldacena2016,Jian2017}.
Especially, the localization of
the protected edge states is not discussed in this Letter.

In our calculation, there exists an intermediate
gapless phase between the non-Abelian topologically ordered phase and the trivial
phase.  This corresponds to the observed crossover between the Kitaev spin liquid
and the Anderson-Kitaev liquid, and the thermal Hall conductivity takes a nonquantized
value in the crossover regime.  In order to
characterize this phase, the longitudinal component $\kappa^{xx}$ is also important~\cite{Garcia2015},
which would be an important future study.  We note that the use of a binary disorder
makes the transition steep, so the calculation by Anderson disorder is also necessary.

As for the computational complexity, a complete $O(N)$ method is possible without
a truncation by a probabilistic approximation of the moments of the Hamiltonian using
multiple random walks.  This new approach was tested and we confirmed its accuracy.
Applying this stochastic $O(N)$ method to our problem setting is also interesting future
work, where essentially $O(10^{5\textrm{--}6})$ sites are achievable.  However,
unfortunately, this approach in the original form only works for the 0-flux sector and
suffers from the ``sign problem'' in other flux sectors, although the problem at hand
is completely classical.  This means that a full $O(N)$ method is not universal and
solving this \textit{classical} sign problem is also a future problem.

In the 0-flux calculation the thermal fluctuation to destroy the
quantization is underestimated, so we need an unbiased Monte Carlo simulation~\cite{Nasu2014,Nasu2017}
to go to finite temperature correctly.  From the previous study, the thermal
Hall conductivity is quantized only in the regime $T/J < O(0.01)$~\cite{Nasu2017}, and this
suggests that the quantization is very weak under bond disorder as well as with thermal
fluctuation.  The combined effect is not sought in this Letter, and thus Monte
Carlo simulations are worth doing.
However, in three dimensions non-Abelian topological phase could potentially survive
under both bond and flux disorders because the flux fluctuation is frozen at finite
$T_c$ by the second-order transition~\cite{Nasu2014}, although we do not know a good
lattice where the thermal conductivity can be quantized in three dimensions.

\bibliography{suppl}